\documentclass[12pt]{article}
\usepackage{multirow}  
\usepackage{amsmath}
\usepackage{amssymb, latexsym}
\usepackage{amsthm}
\usepackage{caption}
\usepackage{setspace}
\usepackage{xcolor}

\def\no{\noindent}
\allowdisplaybreaks

\usepackage{multirow, booktabs, threeparttable}
\def \S{\boldsymbol{S}}

\def \Y{\boldsymbol{Y}}
\def \U{\boldsymbol{U}}
\def \x{\boldsymbol{x}}
\def \b{\boldsymbol{b}}
\def \g{\boldsymbol{g}}

\def \bfp{\boldsymbol{p}}
\def \bftheta{\boldsymbol{\theta}}
\def \bfalpha{\boldsymbol{\alpha}}
\def \bfbeta{\boldsymbol{\beta}}
\def \EL{\scriptscriptstyle EL}
\def \PEL{\scriptscriptstyle PEL}
\def \N{\scriptscriptstyle N}
\def \A{\scriptscriptstyle A}
\def \B{\scriptscriptstyle B}
\def \IPW{\scriptscriptstyle IPW}
\def \DR{\scriptscriptstyle DR}

\newtheorem{theorem}{Theorem}[section]

\begin{document}

\centerline{\large {\bf Pseudo Empirical Likelihood Inference for Non-Probability Survey Samples}}

\bigskip 

\centerline{Yilin Chen$^1$, \ Pengfei Li$^2$,  J.N.K.  Rao$^3$\ and \ Changbao Wu$^2$}
\bigskip 

\centerline{\small$^1$The Hospital for Sick Children, Toronto, Ontario, Canada}
\centerline{\small$^2$Department of Statistics and Actuarial Science, University of Waterloo, Waterloo, Ontario, Canada}
\centerline{\small$^3$School of Mathematics and Statistics, Carleton University, Ottawa, Ontario, Canada}

\bigskip
\bigskip

\hrule

{\small
\begin{quotation}
\no
In this paper, the authors first provide an overview of two major developments on complex survey data analysis: the empirical likelihood methods and statistical inference with non-probability survey samples, and highlight the important research contributions to the field of survey sampling in general and the two topics in particular by Canadian survey statisticians. The authors then propose new inferential procedures on analyzing non-probability survey samples through the pseudo empirical likelihood approach.
The proposed methods lead to asymptotically equivalent point estimators that have been discussed in the recent literature but possess more desirable features on confidence intervals such as range-respecting and data-driven orientation. Results from a simulation study demonstrate the superiority of the proposed methods in dealing with binary response variables. 

\vspace{0.3cm}

\no
KEY WORDS \ Auxiliary information, confidence interval, design-based inference, doubly robust estimator, joint randomization, empirical likelihood ratio statistic, propensity score, reference probability sample.
\end{quotation}
}

\hrule

\bigskip
\bigskip

\section{Introduction}
\label{intro}

The 50th anniversary of the Statistical Society of Canada (SSC) is an occasion to celebrate the contributions of the SSC and Canadian statisticians to statistical sciences. While Canadian statisticians have made major advances in many areas of statistics, the field of survey sampling and official statistics has seen monumental developments on Canadian soil. In this section, we briefly highlight some major contributions of statisticians from Statistics Canada and Canadian universities to survey sampling theory and methods. 

\subsection{Statistics Canada}

Nathan Keyfitz worked at the Dominion Bureau of Statistics (DBS), now called Statistics Canada, from 1936 to 1959 before accepting professorship at renowned universities. He was elected as a Member of the US National Academy of Sciences in recognition of his outstanding research in mathematical demography. He was instrumental in designing the first Canadian Labour Force Survey (LFS) in 1945. During his tenure at Statistics Canada. he published important papers on changing probabilities of selection over time and yet maintaining maximum overlap (Keyfitz, 1951), simplified variance estimation for stratified multistage designs with two primary sampling units per stratum (Keyfitz, 1957) and double ratio estimation, all motivated by practical issues he encountered in the context of LFS.

Ivan Fellegi took charge of survey sampling research soon after Keyfitz left DBS and made major contributions to survey sampling theory and methods. His path-breaking papers on a unified theory for automated record linkage (Fellegi \& Sunter, 1969) and a systematic approach to automatic edit and imputation (Fellegi \& Holt, 1976) were listed among the 19 landmark papers in survey statistics (Jubilee Commemorative Volume, International Association of Survey Statisticians). Fellegi served as Chief Statistician of Canada from 1985 to 2008 until his retirement. He received the SSC Gold Medal Award in 1997, the Lise Manchester Award in 2016 and became an Honorary Member of SSC in 2008. 

David Binder took leadership role in promoting survey sampling theory and methods at Statistics Canada and was instrumental in establishing a division for analysis of complex survey data which resulted in several important papers related to analysis of survey data taking account of design features such as clustering and unequal selection probabilities. His 1983 paper (Binder, 1983) provided a unified method of linearization variance estimation for estimators derived from design-weighted estimating equations. This paper received wide attention (with more than 1300 Google Scholar citations). Binder acted as Director General for Methodology at Statistics Canada before taking early retirement. Binder received the SSC Award for Impact of Applied and Collaborative Work in 2012. 

M. P. Singh was the founding editor of the journal {\em Survey Methodology} published by Statistics Canada. He served as editor from 1975 for 30 years until his untimely death in 2005. Under his editorship, {\em Survey Methodology} became a leading journal in the field of survey sampling and official statistics. After Singh's death, John Kovar, Mike Hidiroglou and Wesley Yung served as editors and J.-F. Beaumont is the current editor. 

Research methodologists at Statistics Canada made major contributions to important problems in survey sampling theory and methods. Topics studied include optimal stratification and allocation (Lavalee \& Hidiroglou, 1988; Bankier, 1986, 1988), indirect sampling when sampling frame that directly corresponds to the target population is not available (Lavallee, 2007), outliers and robust estimation (Beaumont \& Rivest, 2009), composite estimation for LFS (Singh et al., 2001; Gambino et al., 2001), small area estimation and development of a generalized small area estimation system (Estevao et al., 2012), and analysis of complex survey data (Binder, 1983; Roberts et al., 1987; Rubin-Bleuer \& Kratina, 2005).   Y. You received the 2000 Pierre Robillard Award of SSC for his PhD thesis on small area estimation under the supervision of J. N. K. Rao of Carleton University.    

\subsection{Canadian Universities}

Several researchers associated with Canadian universities have made major contributions to survey sampling theory and methods under probability sampling.  Theoretical foundations  were examined by V. P. Godambe (University of Waterloo), by regarding the sample data as the set of sample units, $\S$,  identified through their labels $i\in \S$  and associated values $y_i$  of a study variable $y$.  Under this setup, Godambe derived two important results: (1) Non-existence of the best linear unbiased estimator of the finite population total even under simple random sampling (Godambe, 1955); and (2) The likelihood function is uninformative in the sense it provides no information on the non-sampled values $y_i,i\notin \S$ (Godambe, 1966). The 1955 paper of Godambe was listed among the 19 landmark papers in survey sampling. Godambe was the SSC Gold Medalist in 1987 and became an Honorary Member of SSC in 2001. 

Mary Thompson (University of Waterloo), in collaboration with Godambe, developed a unified theory of estimating functions for making inference from complex survey data, by making use of survey design weights (Godambe \& Thompson, 1986).  Her book ``{\em Theory of Sample Surveys}'' (Thompson, 1997) provided a thoroughly modern and unified treatment that emphasized the connections between theoretical statistics and the field of survey sampling. Mary Thompson received the SSC Gold Medal Award in 2003, the Lise Manchester Award in 2012, and was elected as Honorary Member of SSC in 2021.  

C. E. Sarndal (University of Montreal) developed a unified theory of model-assisted inference which provides design consistent estimators under a working population model regardless of the assumed model. His book on model-assisted survey sampling theory (Sarndal et al., 1992) is a standard textbook on survey sampling and his 1989 paper on variance estimation under the model-assisted approach was listed among the 19 landmark papers in survey sampling (Sarndal et al., 1989). 

J. N. K. Rao (Carleton University) studied methods for analysis of complex survey data taking account of design features such as clustering, stratification and unequal selection probabilities. His 1981 joint paper with A. J. Scott developed simple corrections to chisquare tests for categorical data using the concept of design effects (Rao \& Scott, 1981) and the Rao-Scott corrections are widely used and incorporated in survey data analysis software packages. Rao's 1981 paper was listed among the 19 landmark papers in survey sampling. Rao also wrote two Wiley books on small area estimation, a topic of current interest due to growing demands for local area statistics (Rao, 2003; Rao \& Molina, 2015). Rao received the SSC Gold Medal Award in 1993 and was elected as Honorary Member of SSC in 2004. 

David Bellhouse (Western University) wrote major papers on analysis of complex survey data, systematic sampling, randomized response, optimal estimation under random permutation models and history of survey sampling. He is among the leading researchers in the history of statistics and probability. Bellhouse was elected as Honorary Member of SSC in 2017. 

Steve Thompson (Simon Fraser University) introduced adaptive sampling to deal with hard-to-reach populations. His Wiley book (Thompson \& Seber, 1996) is a standard reference on this important topic of practical importance. His untimely death in January 2021 is a great loss to survey sampling. 

Louis-Paul Rivest (University of Laval) made important contributions to survey sampling theory including capture-recapture methods for estimating animal populations, copula models for small area estimation, optimal stratification, and estimation in the presence of outliers. Rivest received the SSC Gold Medal Award in 2010. 

Jiahua Chen (University of British Columbia) made major contributions to missing data and imputation (Chen \& Shao, 2000, 2001), asymptotic theory for two-phase sampling and empirical likelihood (detailed in Section 2). He received the SSC Gold Medal Award in 2014 and the CRM-SSC Prize in Statistics in 2005.

Randy Sitter (Simon Fraser University) made important contributions to resampling methods for survey data and a unified model-calibration approach jointly with Changbao Wu of the University of Waterloo (Wu \& Sitter, 2001). His tragic death during a 2007 kayak trip in the prime of his career is a great loss to survey sampling as well as to many other areas in statistics to which Sitter made major contributions. The 2001 paper of Wu and Sitter is widely cited. Wu's contributions to empirical likelihood methods for survey data are presented in Section 2. Both Sitter and Wu received the CRM-SSC Prize in Statistics (Sitter in 2004 and Wu in 2012). 

N. G. N. Prasad (University of Alberta) studied the estimation of mean squared error (MSE) of empirical best linear unbiased predictors in small area estimation. His joint paper with Rao (Prasad \& Rao, 1990) provides second order unbiased estimators of MSE and his results are widely used in practice and led to several extensions. The 1990 paper received close to 1000 Google Scholar citations.

Among the younger researchers in Canadian universities working in survey sampling methodology, David Haziza (University of Ottawa) is making major contributions to missing data and imputation and machine learning methods for survey data. Mahmoud Torabi (Univerwsity of Manitoba) is making major contributions to small area estimation and disease mapping methods. Haziza was the recipient of the CRM-SSC Prize in Statistics in 2018. 

It is truly remarkable that five academic survey sampling researchers (Chen, Godambe, Rao, Rivest and M. Thompson) were awarded the SSC Gold Medal and four of those (Godambe, Rao, Rivest and M. Thompson) were elected as Fellows of the Royal Society of Canada.

\section{Empirical Likelihood for Complex Survey Data} 
\label{pel}

Parametric models and likelihood-based inferences are one of the pillars of modern statistics. Parametric inferential procedures have been developed for a variety of statistical problems, with elegant properties such as optimal estimators and most powerful tests. Maximum likelihood estimators based on the assumed parametric models are usually very efficient and the Wilks' theorem holds for a large class of hypothesis test problems where the likelihood ratio statistic has a standard chisquare limiting distribution. Parametric models are also the foundation for Bayesian inference.  

Survey sampling is an important branch of modern statistical sciences. Surveys are one of the primary data collection tools for many fields including official statistics, social and health sciences, and economic studies. There are two distinct features in survey sampling: the target population is finite and the sampled units are chosen based on a survey design that typically involves stratification, clustering and unequal probability selection. Many statistical procedures developed for independent observations from a conceptual infinite population cannot be used for finite population inferences with complex survey samples because of the aforementioned features of survey data. 

Empirical likelihood was first proposed by Owen (1988) for independent and identically distributed (iid) samples. It is a nonparametric likelihood that possesses many attractive features similar to parametric likelihood. In particular, a nonparametric version of the Wilks' theorem holds for standard hypothesis testing problems. Owen (2001) contains a comprehensive review on empirical likelihood. It turns out that the empirical likelihood is a powerful inferential tool for analysis of complex survey data. The discrete nonparametric likelihood is well suited for finite populations. The survey design features can be partially built into the likelihood function through the use of first order inclusion probabilities, and valid design-based inferences can be achieved through the incorporation of the design effect and/or design-based variance estimation. Rao \& Wu (2009) and Chapter 8 of Wu \& Thompson (2020) present an overview of the empirical likelihood methods for complex survey data. 

\subsection{Empirical Likelihood for iid Samples}
\label{EL-0}

Suppose that $(Y_1,\cdots,Y_n)$ is an iid sample from $Y$ with mean $\mu_0=E(Y)$.  Let $\bfp=(p_1,\cdots,p_n)$ be the discrete probability measure over the $n$ sampled units. The empirical (log) likelihood function is defined as 
\begin{equation}
\ell_{\EL}(\bfp) = \log \left(\prod_{i=1}^np_i\right) = \sum_{i=1}^n \log(p_i)\,.
\label{EL}
\end{equation}
In the absence of any additional information, maximizing the empirical likelihood function $\ell_{\EL}(\bfp)$ subject to the normalization constraint 
\begin{equation}
\sum_{i=1}^n p_i = 1
\label{norm}
\end{equation}
leads to the ``global'' maximizer $\hat{p}_i = n^{-1}$, $i=1,\cdots, n$. Note that the use of the log-likelihood automatically requires $p_i>0$ for all $i$. The parameter of interest, $\mu_0=E(Y)$, leads to the so-called parameter constraint 
\begin{equation}
\sum_{i=1}^n p_i Y_i = \mu
\label{mu}
\end{equation}
for a given $\mu$. Under the normalization constraint (\ref{norm}), the constraint (\ref{mu}) can be equivalently written as $\sum_{i=1}^np_i(Y_i-\mu)=0$, corresponding to the so-called moment condition $E(Y-\mu_0)=0$. 

The first milestone on empirical likelihood is the nonparametric version of the Wilks' theorem established by Owen (1988). Let $\hat{p}_i(\mu)$, $i=1,\cdots,n$ be the ``restricted'' maximizer of $\ell_{\EL}(\bfp)$ subject to both the normalization constraint (\ref{norm}) and the parameter constraint (\ref{mu}) for a given $\mu$. The empirical likelihood ratio statistic is given by $r_{\EL}(\mu) = \ell_{\EL}(\hat{\bfp}(\mu)) - \ell_{\EL}(\hat{\bfp}) = \sum_{i=1}^n\log\{n\hat{p}_i(\mu)\}$. 
Owen (1988) showed that $-2r_{\EL}(\mu_0)$ converges in distribution to a $\chi^2$ random variable with one degree of freedom under suitable finite moment conditions on $Y$.

The second milestone on empirical likelihood is the paper by Qin and Lawless (1994) on combining empirical likelihood with general estimating equations.  It opens the door for empirical likelihood to become a general inferential tool for a variety of statistical problems. Let $\Y$ be vector-valued and $\bftheta$ be a $k\times 1$ vector of parameters. The true values of the parameters, $\bftheta_0$, can often be defined under a semiparametric setting through the so-called moment conditions, $E\{\g(\Y,\bftheta_0)\} = {\bf 0}$, where $\g(\cdot,\cdot)$ is an $r\times 1$ vector of real-valued estimating functions. The moment conditions can naturally be incorporated into the empirical likelihood through sample-based constraints 
\begin{equation}
\sum_{i=1}^np_i \, \g(\Y_i,\bftheta) = {\bf 0}\,.
\label{g}
\end{equation} 
The profile empirical likelihood function is given by $\ell_{\EL}(\hat{\bfp}(\bftheta)) = \sum_{i=1}^n\log(\hat{p}_i(\bftheta))$, where $\hat{p}_i(\bftheta)$, $i=1,\cdots,n$ maximize $\ell_{\EL}(\bfp)$ subject to (\ref{norm}) and (\ref{g}) with the given $\bftheta$. The maximum empirical likelihood estimator of $\bftheta_0$ is defined as the maximizer of $\ell_{\EL}(\hat{\bfp}(\bftheta))$. The empirical likelihood ratio statistic is defined as $r_{\EL}(\bftheta) = \ell_{\EL}(\hat{\bfp}(\bftheta)) - \ell_{\EL}(\hat{\bfp}) = \sum_{i=1}^n\log\{n\hat{p}_i(\bftheta)\}$.
A multivariate nonparametric Wilks' theorem was established by Qin and Lawless (1994). 

The general form of the constraints (\ref{g}) includes many important scenarios for practical applications. The first major scenario is when the estimating functions are just-identified, i.e., $r=k$, the number of equations in (\ref{g}) equals to the number of parameters. The maximum empirical likelihood estimator of $\bftheta_0$ is the same as the $m$-estimator (Newey \& McFadden, 1994) that solves the system of equation equations $n^{-1}\sum_{i=1}^n\g(\Y_i,\bftheta) = {\bf 0}$. The second major scenario is when the estimating equations system is over-identified, i.e., $r>k$, the number of equations in (\ref{g}) is larger than the number of parameters. This can happen, for instance, when a single parameter satisfies more than one moment condition.  One such example is the rate parameter from a Poisson distribution where the mean and the variance are the same. Another important practical scenario for over-identified estimation equations system is the availability of auxiliary population information, which can be used to form additional constraints. The constraints (\ref{g}) also allow the estimating functions $\g(\Y,\bftheta)$ to be non-smooth in $\bftheta$, which covers a large class of inference problems such as quantile estimation and quantile regression analysis.

\subsection{Pseudo Empirical Likelihood and Sample Empirical Likelihood}
\label{EL-1}

As noted in Section 1, Godambe (1966) showed that the likelihood function provides no information on the non-sampled values, under the label-based sample data, regarding the unknown population values $(y_1,\cdots,y_{\N})$ as the parameter vector. Hartley and Rao (1968) proposed an alternative approach that suppresses some aspects of the sample data, depending on the situation, to make the sample non-unique and in turn the likelihood function informative. They called their method the ``{\em scale-load approach}'' and illustrated its use under simple random sampling or stratified simple random sampling. As an example, under stratified simple random sampling the strata labels are retained because of known strata differences but the labels within strata are suppressed in the absence of known information regarding the within strata labels. Hartley and Rao (1968) also showed that the scale-load approach provides an effective method for using known population mean of an auxiliary variable   associated with the study variable $y$. Hartley and Rao (1969) generalized the scale-load approach to unequal probability sampling with replacement where selection probability is proportional to size of the unit. 

The standard empirical likelihood for iid samples was applied directly to survey data by Chen and Qin (1993) under simple random sampling and by Zhong and Rao (2000) for stratified simple random sampling. Under a general probability sampling design and the framework for design-based inference, there are two parallel approaches to survey data analysis: the pseudo empirical likelihood and the sample empirical likelihood. Let $\pi_i=P(i\in \S)$ be the first order inclusion probabilities, where $\S$ is the set of $n$ sampled units. Let $d_i = \pi_i^{-1}$ be the basic design weights. 

The pseudo empirical likelihood for complex survey data was first proposed by Chen and Sitter (1999), with a focus on point estimation using auxiliary population information. The pseudo empirical likelihood function defined in Wu and Rao (2006) is given by 
\begin{equation}
\ell_{\PEL}(\bfp) = n\sum_{i\in \S}\tilde{d}_i(\S)\log(p_i)\,,
\label{PEL}
\end{equation}
where $\tilde{d}_i(\S) = d_i /\sum_{j\in \S}d_j$ are the normalized design weights such that $\sum_{i\in \S}\tilde{d}_i(\S)=1$. The pseudo empirical likelihood function $\ell_{\PEL}(\bfp)$ incorporates the design features through the first order inclusion probabilities. Under simple random sampling, we have $\tilde{d}_i(\S)=n^{-1}$, and $\ell_{\PEL}(\bfp)$ reduces to the standard empirical likelihood function for iid samples. 

The pseudo empirical likelihood function $\ell_{\PEL}(\bfp)$, coupled with the standard normalization constraint (\ref{norm}) and the standard parameter constraints (\ref{g}), leads to valid point estimators under the design-based framework. The ``global'' maximizer of $\ell_{\PEL}(\bfp)$ under the constraint (\ref{norm}) is given by $\hat{p}_i = \tilde{d}_i(\S)$. Let $\hat{p}_i(\bftheta)$, $i\in \S$ be the ``restricted'' maximizer of $\ell_{\PEL}(\bfp)$ under both (\ref{norm}) and (\ref{g}) with the given $\bftheta$. The maximum pseudo empirical likelihood estimator $\hat{\bftheta}$ is the maximizer of the profile function $\ell_{\PEL}(\hat{\bfp}(\bftheta))$. Consider the just-identified scenario where $r=k$. It can be shown that the estimator $\hat{\bftheta}$ is identical to the solution to the survey weighted estimating equations 
\begin{equation}
\U_n(\bftheta) = \sum_{i\in \S} d_i \, \g(\Y_i,\bftheta) = {\bf 0}\,.
\label{gee}
\end{equation} 
Under suitable regularity conditions (Binder, 1983), the estimator $\hat{\bftheta}$ is design-consistent to the census parameters $\bftheta_{\N}$ defined as the solution to the census estimating equations $\U_{\N}(\bftheta) = \sum_{i=1}^N  \g(\Y_i,\bftheta) = {\bf 0}$, where $N$ is the population size. 

The pseudo empirical likelihood ratio statistic $r_{\PEL}(\bftheta) = \ell_{\PEL}(\hat{\bfp}(\bftheta)) -  \ell_{\PEL}(\hat{\bfp})$ does not lead to a nonparametric version of the Wilks' theorem under general unequal probability survey designs. It is apparent that the pseudo empirical likelihood function $\ell_{\PEL}(\bfp)$ only involves the first order inclusion probabilities, and valid design-based confidence intervals and hypothesis tests typically require second order inclusion probabilities $\pi_{ij}=P(i,j\in \S)$. It was shown by Zhao and Wu (2019) that the limiting distribution of $-2r_{\PEL}(\bftheta_{\N})$ is a weighted $\chi^2$ with the weights involving the design-based variance of $\U_n(\bftheta_{\N})$. When $\bftheta$ is a scalar, the limiting distribution of $-2r_{\PEL}(\bftheta_{\N})/{\rm deff}$ is a standard $\chi^2$ with one degree of freedom, where ``deff'' is the design effect (Wu and Rao, 2006). 

The survey design features can also be incorporated through the sample empirical likelihood approach. The standard empirical likelihood function given in (\ref{EL}) remains unchanged but the parameter constraints are replaced by the following survey weighted version
\begin{equation}
\sum_{i\in \S}p_i \{d_i\g(\Y_i,\bftheta)\} = {\bf 0}\,.
\label{gd}
\end{equation} 
The basic design weights $d_i$ are treated as part of the estimating functions. The ``global'' maximizer of $\ell_{\EL}(\bfp)$ is given by $\hat{p}_i=n^{-1}$, $i\in \S$. For the scenario of just-identified estimating equations where $r=k$, the maximum sample empirical likelihood estimator $\hat{\bftheta}$ solves the equations system $\sum_{i\in \S}n^{-1} \{d_i\g(\Y_i,\bftheta)\} = {\bf 0}$, 
which is once again identical to the survey weighted estimating equations estimator given as the solution to (\ref{gee}).

The term ``sample empirical likelihood'' was first used by Chen \& Kim (2014) as a contrast to the so-called population empirical likelihood. The idea of using the inclusion/response probabilities as part of the constraints for valid empirical likelihood inference to achieve inverse probability weighting was first proposed by Qin \& Zhang (2007) for missing data problems. It was also used in empirical likelihood methods for complex survey data by Chen \& Kim (2014), Berger \& Torres (2016) and Oguz \& Berger (2016). The sample empirical likelihood as a general inference tool for survey data analysis, including tests of general hypotheses and design-based variable selection, was developed in the paper by Zhao et al. (2022).

The pseudo empirical likelihood uses a survey weighted version of the empirical likelihood function and standard parameter constraints. The sample empirical likelihood adapts the standard empirical likelihood function but employs a survey weighted version of the constraints. Both approaches lead to valid statistical inference under the design-based framework. Some limited simulation results for comparing the two approaches were reported in Zhao \& Wu (2019). Theoretical comparisons between the pseudo empirical likelihood and the sample empirical likelihood, however, are not available in the existing literature. In this paper, we present the pseudo empirical likelihood methods for non-probability survey samples. Using the sample empirical likelihood for integrating data from a non-probability survey sample and a reference probability survey sample is not studied in the current paper.

\subsection{Contributions of Canadian Statisticians to Empirical Likelihood} 
\label{EL-2}

The empirical likelihood was invented by Art Owen of Stanford University, a Canadian statistician who received the SSC Gold Medal in 2021. In the preface of his 2001 monograph (Owen, 2001) on empirical likelihood, Owen wrote: ``As an undergraduate, I studied statistics and computer science at the University of Waterloo. The statistics professors there instilled in me a habit of turning first to the likelihood function, whenever an inference problem appeared''. In his virtual acceptance speech for the SSC gold medal, Owen stated: ``Without University of Waterloo, the empirical likelihood wouldn't have been invented.''

Jing Qin started working on empirical likelihood when he was a PhD student in statistics at the University of Waterloo under the supervision of Jerry Lawless. The 1994 annals paper (Qin \& Lawless, 1994) on combining empirical likelihood with general estimating equations, as well as several other papers on empirical likelihood, was part of his PhD dissertation. Jing Qin received the 1993 SSC Pierre Robillard Award, and Jerry Lawless was the SSC Gold Medalist in 1999. 

The first paper on pseudo empirical likelihood (Chen \& Sitter, 1999) was by Jiahua Chen of University of British Columbia (formerly University of Waterloo) and the late Randy Sitter of Simon Fraser University. Two of the co-authors of the current paper, J. N. K. Rao and Changbao Wu, have made important contributions to empirical likelihood for complex survey data (Wu \& Rao, 2006; Rao \& Wu, 2010a, 2010b; Zhao et al., 2020a, 2020b). Several recent papers on pseudo empirical likelihood and sample empirical likelihood, including Zhao \& Wu (2019), Zhao et al. (2020a, 2020b, 2020c) and Zhao et al. (2022), were led by Puying Zhao who was a postdoctoral research fellow at the University of Waterloo under the supervision of Changbao Wu and David Haziza of University of Ottawa (formerly University of Montreal). Other Canadian statisticians who made noticeable contributions to empirical likelihood include another co-author of the current paper, Pengfei Li of University of Waterloo, and Min Tsao of University of Victoria.

\section{Non-Probability Survey Samples}
\label{nonprob}

Probability survey designs and design-based inference have been widely used by official statistics and researchers in many fields of scientific investigations. In recent years, however, non-probability survey samples have emerged as a convenient and important data source. The penetration of the Internet into every corners of the society in the past two decades has made web-based surveys a popular tool for researchers using the so-called web-panels. A web-panel is a list of willing-participants constructed and maintained by a research organization or a commercial company. The most prominent issue with survey samples collected through web-panels is that the samples are typically biased with unknown inclusion/participation probabilities. 

The biased nature of non-probability survey samples cannot be corrected using the sample itself. Valid statistical inference requires auxiliary information from the target population. See, for instance, Chen et al. (2020) for further discussions. A popular approach for analyzing non-probability survey samples is developed when  there is an existing probability survey with available information on auxiliary variables. Inverse probability weighting and doubly robust estimation for the finite population mean of the study variable have been developed in the paper by Chen et al. (2020) under this setting. We provide a brief review of the methods in this section, which is part of the foundation for the pseudo empirical likelihood methods to be presented in Section \ref{pelnonprob}.

\subsection{Inverse Probability Weighting}
\label{NP-1}

Let $(y_i,\x_i)$ be the values of the study variable $y$ and the vector $\x$ of auxiliary variables associated with unit $i$ in the finite population, $i=1,\cdots,N$. Let $\mu_y = N^{-1}\sum_{i=1}^Ny_i$ be the finite population mean of $y$. The non-probability survey sample $\S_{\A}$ of size $n_{\A}$ collects information on both $y$ and $\x$, and the dataset is denoted as $\{(\x_i,y_i),i\in \S_{\A}\}$. In addition, there exists a probability sample $\S_{\B}$ of size $n_{\B}$, with information available on the auxiliary variables $\x$ but not on the study variable $y$. The dataset from the probability sample is denoted as $\{(\x_i,d_i^{\B}), i\in \S_{\B}\}$, where the $d_i^{\B}$ are the survey weights and are part of of the existing data file. 

The construction of the inverse probability weighted estimator of $\mu_y$ requires the propensity scores $\pi_i^{\A} = P(i\in \S_{\A}\mid \x_i,y_i)$, $i\in \S_{\A}$. 
The term ``propensity scores'' was borrowed from the literature on missing data analysis and causal inference. It was used by Chen et al. (2020) and Kim et al. (2021), among others. The $\pi_i^{\A}$ were also termed as ``participation probabilities'' by several other authors; see, for instance, Beaumont (2020) and Rao (2021). Unlike probability survey samples where the inclusion probabilities are known from the survey design, the propensity scores are unknown for non-probability samples. Let $\hat{\pi}_i^{\A}$ be a suitable estimate of $\pi_i^{\A}$. The inverse probability weighted (IPW) estimator of $\mu_y$ follows the Horvitz-Thompson estimator used in survey sampling and is given by 
\begin{equation}
\hat{\mu}_{\IPW 1} = \frac{1}{N}\sum_{i\in \S_{\A}}\frac{y_i}{\hat{\pi}_i^{\A}} \;\;\;\;\;\; {\rm or} \;\;\;\;\;\;
\hat{\mu}_{\IPW 2} = \frac{1}{\hat{N}^{\A}}\sum_{i\in \S_{\A}}\frac{y_i}{\hat{\pi}_i^{\A}}\,,
\label{IPW}
\end{equation}
where $\hat{N}^{\A}=\sum_{i\in \S_{\A}}1/\hat{\pi}_i^{\A}$ is the estimated population size and $\hat{\mu}_{\IPW 2}$ corresponds to the H\'{a}jek estimator of $\mu_y$ in survey sampling. 

It turns out that one of the most crucial issues in analyzing non-probability samples is to obtain valid estimates for the propensity scores. 
Let $R_i=I(i\in \S_{\A})$ be the sample inclusion indicator variables, $i=1,\cdots,N$. Chen et al. (2020) developed valid estimation procedures for the  propensity scores under the following three basic assumptions:
\begin{itemize}
\item[\rm] {\bf A1.} The sample inclusion indicator variable $R_i$ and the response variable $y_i$ are independent given the covariates $\x_i$.
\item[\rm] {\bf A2.} All units in the target population have a positive propensity score, i.e., $\pi_i^{\A}>0$ for all $i$. 
\item[\rm] {\bf A3.} The indicator variables $R_i$ and $R_j$ are independent given $\x_i$ and $\x_j$ for $i\ne j$. 
\end{itemize}

Under the assumption {\bf A1} we have
$\pi_i^{\A} = P(R_i = 1 \mid \x_i,y_i) = P(R_i = 1 \mid \x_i)$ and a parametric form of $\pi_i^{\A}$ may be imposed as $\pi_i^{\A} = \pi(\x_i, \bfalpha)$, where $\pi(\cdot,\cdot)$ has a known functional form. Let $\bfalpha_0$ be the true value of the parameters. The full log-likelihood function for $\bfalpha$ is given by 
\begin{equation}
\ell(\bfalpha) = \sum_{i=1}^N\Big\{R_i\log \pi_i^{\A}+(1-R_i)\log(1-\pi_i^{\A}) \Big\}
=\sum_{i\in \S_{\A}}\log\Big(\frac{\pi_i^{\A}}{1-\pi_i^{\A}} \Big) + \sum_{i=1}^N\log(1-\pi_i^{\A}) \,.
\label{ell}
\end{equation}
The likelihood function $\ell(\bfalpha)$ given in (\ref{ell}) is not computable based on the non-probability sample alone. The term  $\sum_{i=1}^N\log(1-\pi_i^{\A})$ requires information on $\x$ at the population level. This is where the existing probability survey sample $\S_{\B}$ is used to replace the population total by 
$\sum_{i\in \S_{\B}}d_i^{\B}\log\{1-\pi(\x_i,\bfalpha)\}$ with any given $\bfalpha$. It can be shown that the resulting maximum pseudo-likelihood estimator $\hat{\bfalpha}$ is consistent for $\bfalpha_0$ under the joint randomization of the model for the propensity scores and the survey design for the probability sample $\S_{\B}$ (Chen et al., 2020). The estimated propensity scores are computed as $\hat{\pi}_i^{\A} = \pi(\x_i,\hat{\bfalpha})$, $i\in \S_{\A}$. 

It is also possible to estimate $\bfalpha_0$ through a calibration-based approach. It can be shown that the following estimating equations system is unbiased in the sense that the expectation of the left hand side of (\ref{ee}) is zero with respect to the assumed model on the propensity scores $\pi(\x_i,\bfalpha)$, 
\begin{equation}
\sum_{i\in \S_{\A}}\frac{\x_i}{\pi(\x_i,\bfalpha)} - \sum_{i=1}^N\x_i = {\bf 0}\,,
\label{ee}
\end{equation}
when $\bfalpha=\bfalpha_0$. In addition to the non-probability survey sample, the equation system (\ref{ee}) only requires the population totals for auxiliary variables, which are sometimes available from existing sources. The estimator $\hat{\bfalpha}$ can be obtained by solving (\ref{ee}). If the population totals $\sum_{i=1}^N\x_i $ are not available, we can replace them by $\sum_{i\in \S_{\B}}d_i^{\B}\x_i$ using the probability survey sample $\S_{\B}$. The calibration-based estimator and the maximum pseudo-likelihood estimator of the model parameters $\bfalpha$, however, are not algebraically equivalent. Both estimators are consistent under the assumed parametric form $\pi_i^{\A} = \pi(\x_i,\bfalpha)$ for the propensity scores. 

\subsection{Doubly Robust Estimation}
\label{NP-2}

Doubly robust estimators of $\mu_y$ are constructed using both the propensity score model and the so-called outcome regression model on $y$ given $\x$, and the resulting estimator is consistent if one of the two models is correctly specified. Let $\hat{\pi}_i^{\A}$, $i\in \S_{\A}$ be the estimated propensity scores. 

Consider a semi-parametric model $E(y\mid \x)=m(\x,\bfbeta)$, $V(y\mid \x) =v(\x)\sigma^2$ for the outcome regression, where $m(\cdot,\cdot)$ and $v(\cdot)$ have known functional forms and $\bfbeta$ and $\sigma^2$ are unknown parameters. The $y_i$'s are also assumed to be independent given the $\x_i$'s. Let $\bfbeta_0$ be the true value of $\bfbeta$. Under the assumption {\bf A1}, i.e., $P(R_i = 1 \mid \x_i,y_i) = P(R_i = 1 \mid \x_i)$, the parameters $\bfbeta_0$ for the outcome regression model can be consistently estimated by using the data from the non-probability sample, $\{(\x_i,y_i),i\in \S_{\A}\}$. 
The weighted least square estimator $\hat{\bfbeta}$ of $\bfbeta_0$ minimizes the weighted sum of squares of the residuals, 
\[
S(\bfbeta) = \sum_{i\in \S_{\A}} \{v(\x_i)\}^{-1}\big\{y_i - m(\x_i,\bfbeta)\}^2\,.
\]
Let $\hat{m}_i=m(\x_i,\hat{\bfbeta})$ be the fitted values from the outcome regression model. If the complete auxiliary information $\{\x_1,\cdots,\x_{\N}\}$ is available, the $\hat{m}_i$ can be computed for $i=1,\cdots,N$. The standard doubly robust estimator of $\mu_y$ is then constructed as 
\[
\hat{\mu}_{\DR} = \frac{1}{N}\sum_{i\in \S_{\A}}\frac{y_i-\hat{m}_i}{\hat{\pi}_i^{\A}} +\frac{1}{N}\sum_{i=1}^N\hat{m}_i\,.
\]

A practically important scenario is when the outcome regression model is linear and $m(\x,\bfbeta)=\x'\bfbeta$. In this case the term 
$N^{-1}\sum_{i=1}^N\hat{m}_i$ reduces to $\mu_{\x}'\hat{\bfbeta}$, where $\mu_{\x}=N^{-1}\sum_{=1}^N\x_i$ is the vector of population means of the auxiliary variables. Under the setting considered in this paper with the existence of a probability sample on auxiliary information, Chen et al. (2020) proposed the following two versions of the doubly robust estimator of $\mu_y$:
\begin{eqnarray}
\hat{\mu}_{\DR 1} &=& \frac{1}{N}\sum_{i\in \S_{\A}}\frac{y_i-\hat{m}_i}{\hat{\pi}_i^{\A}} +\frac{1}{N}\sum_{i\in \S{\B}}d_i^{\B}\hat{m}_i \,, 
\label{DR1}\\
\hat{\mu}_{\DR 2} &=& \frac{1}{\hat{N}^{\A}}\sum_{i\in \S_{\A}}\frac{y_i-\hat{m}_i}{\hat{\pi}_i^{\A}} +\frac{1}{\hat{N}^{\B}}\sum_{i\in \S{\B}}d_i^{\B}\hat{m}_i \,,
\label{DR2}
\end{eqnarray}
where $\hat{N}^{\A}=\sum_{i\in \S_{\A}}1/\hat{\pi}_i^{\A}$ and $\hat{N}^{\B}=\sum_{i\in \S_{\B}}d_i^{\B}$ are two versions of the estimated population size. 
The two estimators given in (\ref{DR1}) and (\ref{DR2}) are doubly robust in the sense that they are consistent for $\mu_y$ if one of the working models for the propensity scores and the outcome regression is correctly  specified. Simulation results showed that the estimator $\hat{\mu}_{\DR 2}$ using the estimated $N$ has better performance in terms of bias, a property that could be attributed to the H\'{a}jek estimator. 

Statistical analysis of non-probability survey samples is part of the larger current research topic on integration of data from multiple sources. Data integration has attracted significant amount of attention in recent years among survey statisticians and official statistical agencies. The results described in Sections \ref{NP-1} and \ref{NP-2} are part of the doctoral dissertation of Yilin Chen (Chen, 2020) completed at the University of Waterloo under the supervision of Pengfei Li and Changbao Wu, two of the co-authors of the current paper. Other Canadian statisticians who are active researchers on the topic include Jean-Francois Beaumont of Statistics Canada (Beaumont, 2020) and another co-author of the current paper, J.N.K. Rao of Carleton University (Rao, 2021).

\section{Pseudo Empirical Likelihood for Non-Probability Samples}
\label{pelnonprob}

The pseudo empirical likelihood approach described in Section \ref{EL-1} can be adapted for inference to non-probability survey samples using the estimated propensity scores $\hat{\pi}_i^{\A} = \pi(\x_i,\hat{\bfalpha})$, $i\in \S_{\A}$. The pseudo empirical likelihood function for the non-probability survey sample $\S_{\A}$ is defined similar to (\ref{PEL}) and is given by
\begin{equation}
\ell_{\PEL}(\bfp) = n_{\A}\sum_{i\in \S_{\A}}\tilde{d}_i^{\A}\log(p_i)\,,
\label{PELA}
\end{equation}
where $\bfp = (p_1,\cdots,p_{n_{\A}})$, $\tilde{d}_i^{\A} = (\hat{\pi}_i^{\A})^{-1} /\hat{N}^{\A}$ and $\hat{N}^{\A} = \sum_{j\in \S_{\A}}(\hat{\pi}_j^{\A})^{-1}$ which is defined earlier in Section \ref{nonprob}. Without using any additional information, maximizing $\ell_{\PEL}(\bfp)$ under the normalization constraint 
\begin{equation}
\sum_{i\in \S_{\A}}p_i = 1
\label{norm2}
\end{equation}
leads to $\hat{p}_i = \tilde{d}_i^{\A}$, $i\in \S_{\A}$. The maximum pseudo empirical likelihood estimator of $\mu_y$ is given by 
$\hat{\mu}_{\PEL} = \sum_{i\in \S_{\A}}\hat{p}_i y_i$, which is identical to the IPW estimator $\hat{\mu}_{\IPW 2}$ given in (\ref{IPW}). 

The pseudo empirical likelihood approach to inference for non-probability survey samples possesses several attractive features. First, the doubly robust estimator of $\mu_y$ can be computed in the form of $\sum_{i\in \S_{\A}}\hat{p}_i y_i$ by incorporating a model-calibration constraint (Wu \& Sitter, 2001) into the maximization process.  Second, the pseudo empirical likelihood ratio confidence intervals or tests can be constructed using a scaled chisquare  limiting distribution (Wu \& Rao, 2006) or a bootstrap calibration procedure (Wu \& Rao, 2010). Third, it allows the use of available auxiliary population information through the inclusion of additional moment constraints. Other features of empirical likelihood such as range-respecting and transformation-invariant also become attractive for non-probability survey samples.

\subsection{Doubly Robust Estimation}
\label{PEL-DR}

In practice, the study variable $y$ for non-probability samples is often binary, and the population mean $\mu_y$ becomes the population proportion with the restricted range $[0,1]$.  Commonly encountered examples with binary responses include public opinion surveys or election polls where non-probability surveys become increasingly popular. The doubly robust estimators $\hat{\mu}_{\DR 1}$ and $\hat{\mu}_{\DR 1}$ given in (\ref{DR1}) and (\ref{DR2}) may not respect the range of the parameters and can be inefficient, especially when the population proportion is very small or very large. The pseudo empirical likelihood approach has clear advantages under such scenarios.

Let $\hat{m}_i = m(\x_i,\hat{\bfbeta})$ be the fitted values under the assumed outcome regression model described in Section \ref{NP-2}. Let $\bar{m}^{\B}=(\hat{N}^{\B})^{-1}\sum_{i\in \S_{\B}}d_i^{\B}\hat{m}_i$ be the estimated population mean of the fitted values using the probability survey sample $\S_{\B}$, where $\hat{N}^{\B} = \sum_{i\in \S_{\B}}d_i^{\B}$. The model-calibration constraint is specified as 
\begin{equation}
\sum_{i\in \S_{\A}}p_i \hat{m}_i = \bar{m}^{\B}\,.
\label{MC}
\end{equation}
Together with the normalization constraint (\ref{norm2}), the model-calibration constraint (\ref{MC}) can be equivalently written as $\sum_{i\in \S_{\A}}p_i (\hat{m}_i - \bar{m}^{\B}) = 0$. Let $\hat{p}_i$, $i\in \S_{\A}$ be the maximizer of $\ell_{\PEL}(\bfp)$ under the two constraints (\ref{norm2}) and (\ref{MC}). It can be shown that $\hat{p}_i = \tilde{d}_i^{\A}/\{1+\lambda (\hat{m}_i-\bar{m}^{\B})\}$, $i\in \S_{\A}$, where the Lagrange multiplier $\lambda$ is the solution to the equation
\[
\sum_{i\in \S_{\A}}\frac{\tilde{d}_i^{\A}(\hat{m}_i-\bar{m}^{\B})}{1+\lambda (\hat{m}_i-\bar{m}^{\B})} = 0\,.
\]
The maximum pseudo empirical likelihood estimator of $\mu_y$ is once again computed as $\hat{\mu}_{\PEL} = \sum_{i\in \S_{\A}}\hat{p}_i y_i$ and 
it is doubly robust in the same spirit as discussed in Section \ref{NP-2}. 

We first argue that the estimator $\hat{\mu}_{\PEL}$ is consistent under the outcome regression model. Note that $\hat{\mu}_{\PEL}$ depends on $\hat{\bfbeta}$ through $\hat{m}_i = m(\x_i,\hat{\bfbeta})$,  and the $\hat{p}_i$'s and the $y_i$'s are not independent given the $\x_i$'s under the outcome regression model. Suppose that we replace $\hat{\bfbeta}$ by $\bfbeta_0$, the true value of the parameters, and use $m_i=m(\x_i,\bfbeta_0)$ to construct the model-calibration constraint (\ref{MC}). It follows that the $\hat{p}_i$'s and the $y_i$'s are conditionally independent given the $\x_i$'s under such a hypothetical setting. Without loss of generality, we use $E(\cdot)$ to denote the conditional expectation under the outcome regression model. We have  
\[
E(\hat{\mu}_{\PEL}) = \sum_{i\in \S_{\A}}\hat{p}_i E(y_i) = \sum_{i\in \S_{\A}}\hat{p}_i m_i =  (\hat{N}^{\B})^{-1}\sum_{i\in \S_{\B}}d_i^{\B} m_i \,.
\]
The last step is due to the model-calibration constraint. It follows immediately that $E(\hat{\mu}_{\PEL}) = E(\mu_y)+O_p(n_{\B}^{-1/2})$, where 
$E(\mu_y) = N^{-1}\sum_{i=1}^Nm_i$. Under suitable regularity conditions including the smoothness of $m(\x,\bfbeta)$ with respect to $\bfbeta$ as described in Chen (2020), we have $\hat{\bfbeta} = \bfbeta_0+O_p(n_{\A}^{-1/2})$ and the initial estimator $\hat{\mu}_{\PEL}$ remains consistent under  the constraint (\ref{MC}) where $\bfbeta_0$ is estimated by $\hat{\bfbeta}$. 

We now present a linearized version of the estimator $\hat{\mu}_{\PEL}$ assuming the model for the propensity scores is correctly specified. The consistency of the estimator $\hat{\mu}_{\PEL}$ under the propensity score model is a by-product of Theorem 1 below. Proofs of Theorems 1--3 presented in this paper follow similar arguments in Chen \& Sitter (1999) and Wu \& Rao (2006), with extra steps and regularity conditions in dealing with the impact of $\hat{\bfalpha}$ and $\hat{\bfbeta}$ in $\hat{\pi}_i$ and $\hat{m}_i$. Details can be found in the doctoral dissertation by Chen (2020) and are omitted here to save space. Note that the estimator $\hat{\mu}_{\IPW 2}$ is defined in (\ref{IPW}). 

\begin{theorem}{Theorem 1.}{}
\label{thm1}
Suppose that the propensity score model is correctly specified. Then under the regularity conditions {\bf C1--C6} listed in Chen (2020) the estimator 
$\hat{\mu}_{\PEL}$ has the following asymptotic expansion,
\[
\hat{\mu}_{\PEL}= \hat{\mu}_{\IPW 2} + \big(\bar{m}^{\B}-\hat{\bar{m}}_{\IPW 2} \big)\hat{B}_{m} +o_p\big(n_{\A}^{-1/2}\big)\,,
\]
where $\hat{\bar{m}}_{\IPW 2}=\sum_{i \in \S_{\A}}\tilde{d}_i^{\A} \hat{m}_i$ and $\hat{B}_{m}=\sum_{i \in \S_{\A}}\tilde{d}_i^{\A}(\hat{m}_i-\bar{m}^{\B})y_i/\sum_{i \in \S_{\A}}\tilde{d}_i^{\A}(\hat{m}_i-\bar{m}^{\B})^2$.  
\end{theorem}

It can be shown that 
$\hat{\bar{m}}_{\IPW 2} = \bar{m}^{\B} + O_p(n_{\A}^{-1/2})$ when the propensity score model is correctly specified, which implies that $\hat{\mu}_{\PEL}= \hat{\mu}_{\IPW 2}+O_p(n_{\A}^{-1/2})$ and the estimator $\hat{\mu}_{\PEL}$ is consistent for $\mu_y$ under the assumed model for propensity scores. 

The asymptotic variance of  $\hat{\mu}_{\PEL}$ can also be derived under the propensity score model. The estimator $\hat\bfbeta$ for the outcome regression model is no longer interpreted for $\bfbeta_0$ but the relation $\hat\bfbeta=\bfbeta^*+O_p(n_{\A}^{-1/2})$ holds for some fixed $\bfbeta^*$ even if the outcome regression model is misspecified (White, 1982). Let $m_i^*=m(\x_i,\bfbeta^*)$ and $\bar{m}^* = N^{-1}\sum_{i=1}^Nm_i^*$. 
It can be shown (Chen, 2020) that $Var(\hat{\mu}_{\PEL})=V_{\PEL}+o(n_{\A}^{-1})$, where 
\begin{align*}
V_{\PEL}&= \frac{1}{N^2}\sum_{i=1}^{N}\frac{1-\pi_{i}^{\A}}{\pi_{i}^{\A}}\big(y_i-m_i^*{B}_{m}^*-h_{\N}-\pi_{i}^{\A}\x_i' {\b}\big)^2+ \frac{1}{N^2}V_p\Big(\sum_{i \in \S_{\B}}d_i^{\B}t_i\Big) \,,
\end{align*}
${B}_{m}^*=\big[\sum_{i=1}^N(m_i^*-\bar{m}^*)^2\big]^{-1}\big[\sum_{i=1}^N(m_i^*-\bar{m}^*)y_i\big]$, 
$h_{\N}=N^{-1}\sum_{i=1}^{N}\big(y_i -m_i^*{B}_{m}^*\big)$, 
\[
\b=\Big\{\sum_{i =1}^{N}\pi_{i}^{\A}(1-\pi_{i}^{\A})\x_i\x_i'\Big\}^{-1}\Big\{\sum_{i =1}^{N}(1-\pi_{i}^{\A})(y_i-m_i^*{B}_{m}^*  -h_{\N})\x_i\Big\} \,,
\]
and $t_i=m_i^*{B}_{m}^*+\pi_{i}^{\A}\x_i' {\b}-\bar{m}^*{B}_{m}^*$. The second term with $V_p(\cdot)$ in $V_{\PEL}$ denotes the variance under the probability survey design for the sample $\S_{\B}$. The variance formula $V_{\PEL}$ is required for Theorem 3 to be presented in Section \ref{PEL-CI}.

\subsection{Pseudo Empirical Likelihood Ratio Confidence Intervals}
\label{PEL-CI}

Confidence intervals and hypothesis tests on $\mu_y$ based on the point estimator $\hat{\mu}_{\PEL}$ require a suitable variance estimator. It turns out that variance estimation for doubly robust estimators does not have a straightforward solution. The validity of the point estimator requires only one of the two models for the outcome regression and the propensity scores to be correctly specified, but there is no need to know which one is correct. Derivations of the asymptotic variance formula, however, require the knowledge of the correctly specified model. This issue remains for the pseudo empirical likelihood ratio confidence intervals.

The pseudo empirical likelihood ratio statistic for $\mu_y$ can be defined based on the IPW estimator $\hat{\mu}_{\IPW 2}$, which does not involve the model-calibration constraint (\ref{MC}). Let $\hat{\bfp} = (\hat{p}_1,\cdots,\hat{p}_{n_{\A}})$ be the ``global'' maximizer of $\ell_{\PEL}(\bfp)$ given in (\ref{PELA}) under the normalization constraint (\ref{norm2}); let $\hat{\bfp}(\mu) = (\hat{p}_1(\mu),\cdots,\hat{p}_{n_{\A}}(\mu))$ be the ``restricted'' maximizer of $\ell_{\PEL}(\bfp)$ subject to (\ref{norm2}) and the constraint induced by the parameter of interest, $\mu_y$, which is given by 
\begin{equation}
\sum_{i\in \S_{\A}}p_i y_i = \mu
\label{muy}
\end{equation}
for a given $\mu$. The pseudo empirical (log) likelihood ratio statistic is defined as 
\[
r_{\PEL}^{(1)}(\mu) = \ell_{\PEL}(\hat\bfp(\mu)) - \ell_{\PEL}(\hat\bfp)\,.
\]
The limiting distribution of $r_{\PEL}^{(1)}(\mu_y)$ is a scaled chisquare with one degree of freedom, as shown in Theorem 2 below. The result is similar to Theorem 1 of Wu \& Rao (2006). 

\begin{theorem}{Theorem 2.}{}
\label{thm2}
Suppose that the propensity score model is correctly specified. Then under the regularity conditions {\bf C1--C4} and {\bf C7} listed in Chen (2020), the adjusted pseudo empirical likelihood ratio statistic $-2 r_{\PEL}^{(1)}(\mu)/a_1$ converges in distribution to $\chi^2_1$ when $\mu=\mu_y$. The adjusting factor is computed as $a_1=v_{\IPW}/s_1$, where 
$s_1=n_{\A}^{-1}\sum_{i\in \S_{\A}}\tilde{d}_i^{\A}(y_i-\hat{\mu}_{\IPW 2})^2$ and $v_{\IPW}$ is the variance estimator for $\hat{\mu}_{\IPW 2}$. 
\end{theorem}

The adjusting factor $a_1$ can be viewed as the ``design effect'' for the non-probability survey sample $\S_{\A}$ in terms of the propensity scores. The variance estimator $v_{\IPW}$ is built based on the asymptotic variance formula $V_{\IPW 2}$ given in equation (10) of Chen et al. (2020), which is identical to $V_{\PEL}$ presented in Section \ref{PEL-DR} if we let $B_m^*=0$. Note that $B_m^*=0$ is simply the consequence of a common mean model $E(y_i\mid \x_i)=\beta_0$. The model-calibration constraint (\ref{MC}) under such scenarios reduces to the normalization constraint (\ref{norm2}) and the doubly robust estimator of $\mu_y$ reduces to $\hat{\mu}_{\IPW 2}$. The asymptotic variance consists of two variance components, one associated with the propensity score model and the other with the probability survey design for the reference sample $\S_{\B}$. In practice, a bootstrap calibration procedure to be discussed after Theorem 3 in the next section is more desirable for the implementation of the pseudo empirical likelihood ratio confidence intervals for $\mu_y$. 

The pseudo empirical likelihood ratio statistic for $\mu_y$ can also be defined based on the doubly robust estimator $\hat{\mu}_{\PEL}$ which involves the model-calibration constraint (\ref{MC}). Let $\hat{\bfp} = (\hat{p}_1,\cdots,\hat{p}_{n_{\A}})$ be the ``global'' maximizer of $\ell_{\PEL}(\bfp)$ given in (\ref{PELA}) under the normalization constraint (\ref{norm2}) and the model-calibration constraint (\ref{MC}); let $\hat{\bfp}(\mu) = (\hat{p}_1(\mu),\cdots,\hat{p}_{n_{\A}}(\mu))$ be the ``restricted'' maximizer of $\ell_{\PEL}(\bfp)$ subject to (\ref{norm2}), (\ref{MC}) and the parameter constraint (\ref{muy}) with a given $\mu$. The pseudo empirical (log) likelihood ratio statistic is similarly defined as  $r_{\PEL}^{(2)}(\mu) = \ell_{\PEL}(\hat\bfp(\mu)) - \ell_{\PEL}(\hat\bfp)$.  A result similar to Theorem 2 of Wu \& Rao (2006) can be established under the assumption that the model for the propensity scores is correctly specified. The notations for $\hat{\mu}_{\PEL}$, $\hat{m}_i$, $\bar{m}^{\B}$ and $\hat{B}_m$ follow from Theorem 1 in Section \ref{PEL-DR} on doubly robust estimation.

\begin{theorem}{Theorem 3.}{}
\label{thm3}
Suppose that the propensity score model is correctly specified. Then under the regularity conditions {\bf C1--C7} listed in Chen (2020), the adjusted pseudo empirical likelihood ratio statistic $-2 r_{\PEL}^{(2)}(\mu)/a_2$ converges in distribution to $\chi^2_1$ when $\mu=\mu_y$. The adjusting factor is computed as $a_2=v_{\PEL}/s_2$, where 
$s_2=n_{\A}^{-1}\sum_{i\in \S_{\A}}\tilde{d}_i^{\A}\{y_i-\hat{\mu}_{\PEL} -(\hat{m}_i-\bar{m}^{\B})\hat{B}_m\}^2$ and $v_{\PEL}$ is the variance estimator for $\hat{\mu}_{\PEL}$ under the assumed propensity score model. 
\end{theorem}

The variance estimator $v_{\PEL}$ is built based on the asymptotic variance formula $V_{\PEL}$ presented in Section \ref{PEL-DR}. The $(1-\alpha)$-level pseudo empirical likelihood ratio confidence interval for $\mu_y$ is constructed as $\{\mu \mid -2 r_{\PEL}^{(2)}(\mu)/a_2 \le \chi^2_1(1-\alpha)\}$, where $\chi^2_1(1-\alpha)$ is the $100(1-\alpha)$th quantile from the $\chi^2_1$ distribution. 

Constructions of confidence intervals or hypothesis tests using Theorems 2 and 3 require the adjusting factor $a_1$ and $a_2$, which involves variance estimation for $\hat{\mu}_{\PEL}$. One restrictive feature is that 
the second component in $V_{\PEL}$ depends on the probability sampling design for the sample $\S_{\B}$. A more restrictive feature of the variance formula  is that the formula for $V_{\PEL}$ is derived under the assumed propensity score model. It is invalid when the outcome regression model is correctly specified but the propensity score model is misspecified. This motivates the use of a bootstrap calibration procedure similar to the one described in Wu \& Rao (2010) without involving the adjusting factors $a_1$ or $a_2$. 

The bootstrap calibrated pseudo empirical likelihood ratio confidence interval is constructed as $\{\mu \mid -2 r_{\PEL}(\mu) \le b_{\alpha}\}$, where $r_{\PEL}(\mu)$  is either $r_{\PEL}^{(1)}(\mu)$ or $r_{\PEL}^{(2)}(\mu)$, depending on whether the model-calibration constraint (\ref{MC}) is included, and $b_{\alpha}$ is the upper $100\alpha$th quantile of the sampling distribution of $-2 r_{\PEL}(\mu_y)$ and is approximated through the following bootstrap procedure. 
\begin{itemize}
\item[1.] Select a bootstrap sample $\S_{\A}^{(k)}$ of size $n_{\A}$ from $\S_{\A}$ and a bootstrap sample $\S_{\B}^{(k)}$ of size $n_{\B}$ from $\S_{\B}$ by simple random sampling with replacement. 
\item[2.] Compute the value of $-2 r_{\PEL}(\mu)$ using the bootstrap sample datasets $\{(\x_i,y_i),i\in \S_{\A}^{(k)}\}$ and $\{(\x_i,d_i^{\B}),i \in \S_{\B}^{(k)}\}$ at $\mu=\hat{\mu}_{\PEL}$; denote the value as $b^{(k)}$. 
\item[3.] Repeat Steps 1 and 2 for $k=1,2,\cdots,K$ for a large $K$, independently, to obtain a sequence of values $b^{(1)},b^{(2)},\cdots,b^{(K)}$. Let $b_{\alpha}$ be the $100(1-\alpha)$th sample quantile of the sequence. 
\end{itemize}

Our simulation studies show that the standard with-replacement bootstrap for the probability sample $\S_{\B}$ works well if the survey design is single-stage unequal probability sampling with small sampling fractions. More advanced bootstrap procedures are required if the design involves stratification, clustering or multi-stage unequal probability selections.

\section{Simulation Studies}
\label{simu}

We report results from a simulation study on the performances of confidence intervals on $\mu_y$ where $y$ is a binary variable and $\mu_y$ is the finite population proportion. We consider a finite population of size $N=10,000$ with measurements on $y$, $x_1$, $x_2$ and $x_3$. The outcome regression model is specified as 
\[
\log\Big(\frac{\mu_i}{1-\mu_i} \Big) = \beta_0+\beta_1x_{1i}+\beta_2x_{2i}+\beta_3x_{3i}\,, \;\; i=1,\cdots,N\,,
\]
where $\mu_i = E(y_i\mid \x_i) = P(y_i=1\mid \x_i)$  and $\x_i=(x_{1i},x_{2i},x_{3i})'$. The $\x_i$'s for the finite population are generated as follows: 
$x_{1i}=z_{1i}$, $x_{2i}=z_{2i}+0.1x_{1i}$, $x_{3i}=z_{3i}+0.1x_{2i}$, with $z_{1i}\sim Bernoulli(0.5)$, $z_{2i}\sim Uniform(0,1)$ and $z_{3i}\sim Exp(0.5)$. The $y_i$ given $\x_i$ is generated from $Bernoulli(\mu_i)$ for each $i$. Different values of $\bfbeta=(\beta_0,\beta_1,\beta_2,\beta_3)'$ are considered, which leads to different population proportions $\mu_y$. The results reported in Table \ref{CI2} correspond to $\bfbeta=(-4.1,1.0,1.0,1.0)'$ and $\mu_y=0.1$. Additional simulation results can be found in Chen (2020). 

\begin{table}[!htbp]
\small
\centering
\caption {Performance of Confidence Intervals for the Finite Population Proportion $\mu_y=0.1$}  
\label{CI2} 
\vskip .1cm
\centerline{\tabcolsep=3truept\begin{tabular}{lccrrrrrrrr}
\hline\hline
$(n_{\A},\;n_{\B})$& $\;$ Model $\;$ 	& $\;\;\;\;\;\;\;\;\;\;$ 	&	$PEL_{1, adj}$ 	&	$PEL_{1, bts}$ 	&	$PEL_{2, adj}$	&	$PEL_{2, bts}$	&	$\;\;\;\;$ $NA_{1}$ 	&	$\;\;$ $NA_{2}$	& $\;\;$ $Bst$\\
	\hline
$(100, 200)$ 	&TT	&$\%$CP 	&	92.35	&	92.80	&	91.85&	93.35	&	89.90	&	90.55&88.65\\
	 &&	$\%$L 	&	1.30	&	1.05	&	1.20	&	0.75	&	0.60	&	0.35	&	0.10\\
	&&	$\%$U 	&	6.35	&	6.15	&	6.95	&	5.90	&	9.50	&	9.10	&	11.25\\
	&&	AL 	&	0.127	&	0.131	&	0.124	&	0.132	&	0.128	&	0.130	&	0.128\\
	&FT&	$\%$CP 	&	90.85	&	91.80	&	90.95	&	92.30	&	88.85	&	91.20	&	88.45\\
	&&	$\%$L 	&	2.10&	1.60	&	2.25	&	1.50	&	0.75	&	0.55	&	0.25\\
	&&	$\%$U 	&	7.05	&	6.60	&	6.80	&	6.20	&	10.40	&	8.25	&	11.30\\
	&&	AL 	&	0.126	&	0.130	&	0.125	&	0.130	&	0.127	&	0.132	&	0.131\\
	&TF&	$\%$CP 	&	72.25	&	74.40	&	90.20	&	94.15	&	81.50	&	92.10	&	88.30\\
	&&	$\%$L 	&	27.35	&	25.20	&	2.85	&	0.95	&	18.05	&	0.40	&	0.25\\
	&&	$\%$U 	&	0.40	&	0.40	&	6.95	&	4.90	&	0.45	&	7.50	&	11.45\\
	&&	AL 	&	0.149	&	0.153	&	0.117	&	0.139	&	0.150	&	0.125	&	0.124\\
$(200,100)$ 	&TT&	$\%$CP 	&	93.85	&	94.80	&	92.90&	94.00	&	91.60	&	92.55	&	90.50\\
	&&	$\%$L 	&	1.85	&	1.40	&	1.90	&	1.45	&	0.70	&	0.80	&	0.55\\
	&&	$\%$U 	&	4.30	&	3.80	&	5.20	&	4.55	&	7.70	&	6.65	&	8.95\\
	&&	AL 	&	0.098	&	0.104	&	0.096	&	0.099	&	0.098	&	0.099	&	0.098\\
	&FT&	$\%$CP 	&	93.70	&	94.65	&	93.50	&	93.95	&	91.55	&	93.20	&	91.30\\
	&&	$\%$L 	&	1.85	&	1.40	&	1.85	&	1.55	&	0.65	&	0.90	&	0.50\\
	&&	$\%$U 	&	4.45	&	3.95	&	4.65	&	4.50	&	7.80	&	5.90	&	8.20\\
	&&	AL 	&	0.098	&	0.104	&	0.097	&	0.099	&	0.099	&	0.101	&	0.101\\
	&TF&	$\%$CP 	&	56.25	&	59.70	&	90.40	&	94.90	&	66.15	&	93.65	&	91.25\\
	&&	$\%$L 	&	43.75	&	40.30	&	3.80	&	1.70	&	33.85	&	0.95	&	0.65\\
	&&	$\%$U 	&	0.00	&	0.00	&	5.80	&	3.40	&	0.00	&	5.40	&	8.10\\
	&&	AL 	&	0.112	&	0.117	&	0.088	&	0.102	&	0.113	&	0.094	&	0.094\\
$(200,200)$ 	&TT&	$\%$CP 	&	93.20	&	93.65	&	93.40	&	94.00	&	91.10	&	92.65	&	91.05\\
	&&	$\%$L 	&	1.55	&	1.40	&	1.50	&	1.35	&	1.00	&	0.40	&	0.30\\
	&&	$\%$U 	&	5.25	&	4.95	&	5.10	&	4.65	&	7.90	&	6.95	&	8.65\\
	&&	AL 	&	0.093	&	0.096	&	0.091	&	0.093	&	0.093	&	0.093	&	0.093\\
	&FT&	$\%$CP 	&	92.25	&	93.15	&	92.20	&	92.75	&	90.85	&	92.10	&	90.70\\
	&&	$\%$L 	&	1.65	&	1.35	&	1.50	&	1.35	&	0.80	&	0.70	&	0.40\\
	&&	$\%$U 	&	6.10	&	5.50	&	6.30	&	5.90	&	8.35	&	7.20	&	8.90\\
	&&	AL 	&	0.093	&	0.096	&	0.091	&	0.093	&	0.093	&	0.095	&	0.094\\
	&TF&	$\%$CP 	&	53.00	&	55.35	&	91.05	&	94.25	&	62.55	&	92.95	&	91.65\\
	&&	$\%$L 	&	47.00	&	44.65	&	3.05	&	1.75	&	37.45	&	0.70	&	0.50\\
	&&	$\%$U 	&	0.00	&	0.00	&	5.90	&	4.00	&	0.00	&	6.35	&	7.85\\
	&&	AL 	&	0.108	&	0.111	&	0.086	&	0.097	&	0.108	&	0.089	&	0.089\\
  \hline\hline
 \end{tabular}}
\end{table}

The propensity scores $\pi_i^{\A}$ for the non-probability survey sample follow a logistic regression model 
\[
\log\Big(\frac{\pi_i^{\A}}{1-\pi_i^{\A}} \Big) = \alpha_0+\alpha_1x_{1i}+\alpha_2x_{2i}+\alpha_3x_{3i}\,, \;\; i=1,\cdots,N\,,
\]
with $\alpha_1=\alpha_2=\alpha_3=1$ and a chosen $\alpha_0$ such that $\sum_{i=1}^N\pi_i^{\A}=n_{\A}$ for the given $n_{\A}$. The non-probability sample $\S_{\A}$ is selected by the Poisson sampling method with inclusion probabilities $\pi_i^{\A}$ and the expected sample size $n_{\A}$. The reference probability sample $\S_{\B}$ of size $n_{\B}$ is selected by the Rao-Sampford sampling method (Rao, 1965; Sampford, 1967) with inclusion probabilities $\pi_i^{\B}$ proportional to $z_i=c+x_{3i}+0.03y_i$. The value of $c$ is chosen to control the variation of the survey weights such that $\max z_i / \min z_i = 20$. Three combinations of the sample sizes are considered in Table \ref{CI2}: $(n_{\A},n_{\B})=(100,200)$, $(200,100)$ and $(200,200)$. 

The simulation study is conducted under three different scenarios on model specifications: (i) TT; (ii) FT; and (iii) TF. The first T or F indicates whether the outcome regression model is correctly specified (T) or misspecified (F), and the second T or F represents whether the propensity score model is correctly specified (T) or misspecified (F). The misspecified outcome regression model is given by $\log\{\mu_i/(1-\mu_i)\} = \beta_0+\beta_1x_{1i}+\beta_2x_{2i}$, with the term on $x_{3i}$ missing, and the misspecified propensity score model uses $\log\{\pi_i^{\A}/(1-\pi_i^{\A})\} = \alpha_0+ \alpha_1x_{1i}+\alpha_2x_{2i}$ without the term on $x_{3i}$. 

Seven different methods are used in the simulation to construct a $95\%$ confidence interval on $\mu_y$: The pseudo empirical likelihood ratio confidence interval based on Theorem 2 without the model-calibration constraint (\ref{MC}) using the adjusted $\chi^2_1$ limiting distribution ($PEL_{1,adj}$) or using the bootstrap calibration method ($PEL_{1,bst}$); The pseudo empirical likelihood ratio confidence interval based on Theorem 3 under the model-calibration constraint (\ref{MC}) using the adjusted $\chi^2_1$ limiting distribution ($PEL_{2,adj}$) or using the bootstrap calibration method ($PEL_{2,bst}$); The normal approximation confidence interval to the Wald-statistic based on $\hat{\mu}_{\IPW 2}$ and $v_{\IPW}$ ($NA_1$) or based on $\hat{\mu}_{\DR 2}$ and the bootstrap variance estimator ($NA_2$); The bootstrap percentile confidence interval based on $\hat{\mu}_{\DR 2}$ ($Bst$). 

Performances of confidence intervals are evaluated through the simulated coverage probability (\%CP), the lower tail error rate (\%L), the upper tail error rate (\%U), all in percentages, and the average length (AL). The target value of \%CP is $95$ and we have \%CP$+$\%L$+$\%U$=100$ for all cases. Simulation results are reported in Table \ref{CI2}. Major observations from the simulation study can be summarized as follows: (i) When both models are correctly specified (i.e., TT), all four pseudo empirical likelihood ratio confidence intervals have acceptable performances in terms of coverage probabilities, and are better than the normal approximation based intervals or the bootstrap percentile interval; (ii) When the outcome regression model is misspecified but the propensity score model is correctly specified (i.e., FT), all methods remain valid but with slightly decreased coverage probabilities; (iii) When the outcome regression model is correctly specified but the propensity score model is misspecified (i.e., TF), the intervals without using the model-calibration constraint (\ref{MC}) fail completely and those under the constraint (\ref{MC}) remain valid; (iv) The pseudo empirical likelihood ratio confidence interval with the bootstrap procedure (i.e., $PEL_{2,bst}$) has reliable performance under all scenarios; (v) The sample size $n_{\A}$ for the non-probability sample $\S_{\A}$ plays a more important role for all intervals than the sample size $n_{\B}$ for the reference probability sample $\S_{\B}$.

\section{Concluding Remarks}
\label{conclude}

Statistical analysis of non-probability survey samples requires assumptions on the sample inclusion or participation mechanisms and auxiliary population information. The current setting of two samples, the non-probability sample with measurements on the study variable $y$ and the auxiliary variables $\x$ and an existing reference probability sample with information on the auxiliary variables $\x$, has been used by several authors to develop valid statistical inference procedures. Practical applications of the methods require careful checking of the assumptions and the quality of the auxiliary variables in characterizing the participation behaviour for the non-probability sample as well as the prediction power to the study variable. The pseudo empirical likelihood approach to analyzing non-probability survey samples has certain advantages over other methods as demonstrated in the simulation studies and has potentials to be extended to other directions. One possible extension is the multiply robust inference procedures as discussed in Chen \& Haziza (2017). 

As noted at the end of Section 3, the results presented in the current paper are part of the broader topic on integrating data from different sources. Combining data from two probability survey samples were studied extensively in the existing literature; see, for instance, Wu (2004) and Kim \& Rao (2012) and references therein. The emergence and the increased popularity of non-probability survey samples present both challenges and opportunities to survey statisticians to develop valid and efficient inference procedures under practically feasible settings.

\end{document}